\documentclass[aps,nofootinbib,showpacs,preprintnumbers,amsmath,amssymb]{revtex4}
\usepackage{epsfig}

\begin{document}
\preprint{USM-TH-145, hep-ph/0309262 (v3)}

\title{Estimate of the three-loop contribution to the QCD static potential
from renormalon cancellation}

\author{Gorazd Cveti\v{c}}
  \email{gorazd.cvetic@usm.cl}
\affiliation{Dept.~of Physics, Universidad T\'ecnica
Federico Santa Mar\'{\i}a, Valpara\'{\i}so, Chile}

\date{\today}

\begin{abstract}
It had been known that the Borel transforms of the twice 
quark pole mass $2m_q$ and of the QCD $q$-${\bar q}$ 
static potential $V(r)$ have leading infrared renormalons 
at $b=1/2$ such that they cancel in the sum.
The renormalon residue of $2 m_q$ had been determined with 
reasonably high precision from the known perturbative coefficients 
of the ratio $m_q/{\overline m}_q$, where ${\overline m}_q$ is the 
${\overline {\rm MS}}$ renormalon-free mass of the quark $q$. 
These values of the residues are used here to estimate the 
hitherto unknown part of the three-loop coefficient of the 
static potential $V(r)$. Further, the method takes into
account the fact that the ultrasoft energy regime contributions
must be excluded from the analysis. In the $b {\bar b}$ quarkonium, 
aforementioned estimated part of the three-loop term turns out 
to give a contribution to the binding energy comparable to the 
nonpertubative contribution.

\vspace{1.cm}

\noindent
This is the version v3 as it will appear in J.~Phys.~G.
The changes in comparison to the previous version:
the first paragraph is new; the paragraph containing
Eqs.~(31)-(32) is new, as is Table I; the last paragraph 
(before Acknowledgments) is new; some new references were added. 
Numerical results are unchanged.

\end{abstract}

\maketitle

The physics of heavy quarkonia is an interesting area
of QCD because these systems are expected to be described
to a significant degree, but not entirely, by results of 
perturbative QCD (pQCD). While the perturbation theory
cannot explain the binding phenomenon,
the values of the binding energies of the ground state and 
of some excited states of $b {\bar b}$ and $t {\bar t}$
are expected to be predicted to a reasonable degree by pQCD,
but those of $c {\bar c}$ less well.
Since masses of some of the heavy quarkonia
($b {\bar b}$ and $c{\bar c}$) states
are well measured, their values can be
confronted with predictions based on pQCD,
where estimates of the nonperturbative corrections
should also be included.
Such analyses can give us information on the mass
of the constituents (e.g., of the heavy quark $b$) and predict
masses of other excited states of quarkonium.
One of the important quantities used in such analyses
is the $q$-${\bar q}$ static potential $V(r)$.
The {\em perturbative} part of $V(r)$
is the inverted Fourier transform of the scattering
amplitude for the slowly moving $q {\bar q}$ pair
in the static limit $m_q \to \infty$.
In general, however, $V(r)$ is not reducible to a 
perturbative scattering amplitude that conserves all
components of the total momentum of the two incoming 
and outgoing fermions on their mass-shells.  
In this work we will estimate the unknown part 
of the three-loop coefficient of the 
perturbative static potential,
using the known property 
\cite{Hoang:1998nz,Brambilla:1999xf,Beneke:1998rk}
that the leading infrared renormalon
singularities cancel in the sum $V(r) + 2 m_q$, where
$m_q$ is the pole mass of the quark.

The perturbative expansion of the QCD $q$-${\bar q}$ static potential 
is presently known to two loops 
\cite{Peter:1996ig,Schroder:1998vy}.
In the three-momentum ${\bf k}$-space, 
where $k = (0, {\bf k})$ and ${\bf k} = {\bf p}^{\prime} - {\bf p}$
is the three-momentum transfer between the quark and antiquark,
the potential is written as 
\cite{Schroder:1998vy,Kniehl:2002br,Penin:2002zv}
\begin{eqnarray}
\lefteqn{
V(|{\bf k}|) =  - \frac{16 \pi^2}{3} \frac{1}{|{\bf k}|^2}
a(\mu) {\Bigg \{} 1 + a(\mu) \left[ \frac{1}{4} a_1 + \beta_0 L \right] 
 + a(\mu)^2 \left[ \frac{1}{4^2} a_2 + 
\left( \frac{1}{2} a_1 \beta_0 + \beta_1 \right) L +
\beta_0^2 L^2 \right] } 
\nonumber\\
&& + a(\mu)^3 \left[ \frac{1}{4^3} a_3 + 
b_3 \ln \left( \frac{\mu_f^2}{|{\bf k}|^2} \right)
+ \left( \frac{3}{16} a_2 \beta_0\!+\!\frac{1}{2} a_1 \beta_1 
\!+\!\beta_2 \right) L + 
\left( \frac{3}{4} a_1 \beta_0^2\!+\! 
\frac{5}{2} \beta_0 \beta_1 \right) L^2
+ \beta_0^2 L^3 \right] 
+ {\cal O}(a^4) {\Bigg \}} \ .
\label{Vk}
\end{eqnarray}
Here, $a(\mu) = \alpha_s(\mu,{\overline {\rm MS}})/\pi$; 
$L = \ln(\mu^2/|{\bf k}|^2)$ where $\mu$ is the
renormalization scale;
$\beta_j$ ($j = 0,1,2, \ldots$) are the beta-coefficients
appearing in the renormalization group equation (RGE)
\begin{equation}
\frac{d a(\mu)}{d \ln \mu^2 } = - \beta_0 a^2(\mu) - \beta_1 a^3(\mu)
- \beta_2 a^4(\mu) - \ldots
\label{RGE} 
\end{equation}
We have $\beta_0 = (11 - 2 n_f/3)/4$, $\beta_1 = (102 - 38 n_f/3)/16$.
The higher $\beta_j$ ($j \geq 2$) are renormalization scheme dependent,
and are here taken in ${\overline {\rm MS}}$ scheme.
Here, $n_f\!=\!n_{\ell}$ is the number of the light quark flavors active
in the soft/potential scale regime $|k^2|^{1/2} \sim m_q \alpha_s$
($n_f$ is four for $b$-$\bar b$, five for $t$-$\bar t$).
The terms involving powers of $L$ in Eq.~(\ref{Vk}) are fixed
by the renormalization scale independence of $V$.
The one- and two-loop coefficients $a_1$ and $a_2$ were obtained in
\cite{Fischler:1977yf} and  \cite{Schroder:1998vy}, respectively
\begin{eqnarray}
a_1 &=& \frac{1}{9} ( 93 - 10 n_f) \ ,
\label{a1}
\\
a_2  &=&  456.749 - 66.3542 n_f + 1.23457 n_f^2 \ .
\label{a2}
\end{eqnarray}
At the three-loop level [terms $\sim\!a^4$ in $V(|{\bf k}|)$
of Eq.~(\ref{Vk})],
the coefficient includes the unknown term $a_3$ and a
term with the infrared (IR) cutoff $\mu_f$ 
which cuts out the ultrasoft (US) region 
($|k^2|^{1/2} \sim m_q \alpha_s^2 \sim E_{\rm US}$)
from the soft/potential (S) region 
($|k^2|^{1/2} \sim m_q \alpha_s \sim E_{\rm S}$):
$E_{\rm S} < \mu_f < E_{\rm US}$. The existence of the
IR divergent terms at $\sim\!a^4$ in the static Wilson loop
was pointed out in Ref.~\cite{Appelquist:es}. 
The IR cutoff term 
$b_3 \ln ( \mu_f^2/|{\bf k}|^2 )$ in Eq.~(\ref{Vk}) has
$b_3 = 9 \pi^2/8$, according to Ref.~\cite{Brambilla:1999qa},
and is obtained when a Green function involving color-singlet 
wave function of $q$-${\bar q}$ is matched in two
adjacent effective theories: nonrelativistic QCD (NRQCD)
\cite{Caswell:1985ui} 
where the hard scales ($\sim\!m_q$) are integrated out,
and the potential NRQCD (pNRQCD) \cite{Pineda:1997bj} 
where the soft scales ($\sim\!\alpha_s m_q$) 
are integrated out.\footnote{
In Ref.~\cite{Kniehl:2002br}, the authors calculated the
${\rm N}^3 {\rm LO}$ Hamilton of pNRQCD
(using an estimate for $a_3$ of Ref.~\cite{Chishtie:2001mf})
using a method called threshold expansion where all
the loop integrations 
were performed in $(3\!-\!2 \epsilon)$ dimensions.
This generated some additional non-physical terms in the
ultrasoft and in the soft/potential regime, but those terms
canceled out in the sum from both regimes.}

The three-dimensional Fourier transformation of expansion
(\ref{Vk}) gives the perturbative expansion of the 
static potential in the position-space
\begin{eqnarray}
V(r) & = & - \frac{4 \pi}{3} \frac{1}{r} a(\mu) 
{\Bigg \{} 1 +  a(\mu) \left[ \frac{1}{4} a_1 + 2 \beta_0 {\ell} \right]
\nonumber\\
&&+ a(\mu)^2 \left[ \frac{1}{16} a_2 + 
\left( a_1 \beta_0 + 2 \beta_1 \right) {\ell} +
\beta_0^2 \left( 4 {\ell}^2 + \pi^2/3 \right) \right]
\nonumber\\
&&+ a(\mu)^3  {\bigg [} \frac{1}{4^3} a_3 
+ 2 b_3 \ln( \mu_f r e^{\gamma_{\rm E}}) +
\left( \frac{3}{16} \beta_0 a_2 + \frac{1}{2} \beta_1 a_1 + \beta_2 \right)
2 {\ell} 
\nonumber\\
&&+ \left( \frac{3}{4} \beta_0^2 a_1 + \frac{5}{2} \beta_0 \beta_1
\right) \left( 4 {\ell}^2 + \pi^2/3 \right) 
+\beta_0^3 \left( 8 {\ell}^3 + 2 \pi^2 {\ell} + 16 \xi(3) \right)
{\bigg ]} + {\cal O}(a^4) {\Bigg \}} ,
\label{Vr}
\end{eqnarray}
where we use the notation ${\ell} = \ln[\mu r \exp(\gamma_{\rm E})]$,
with $\gamma_{\rm E}$ the 
Euler constant ($\gamma_{\rm E} = 0.5772 \ldots$).

The term $a_3$ is the only part of the three-loop coefficient
that has not been calculated in the literature.
It is the value of $a_3$ that will be estimated by
using the aforementioned known property that the leading 
infrared renormalon ($b=1/2$) singularities in the sum
$V(r) + 2 m_q$ cancel, $m_q$ being the pole mass of the quark.

\begin{figure}[htb]
\begin{minipage}[b]{.49\linewidth}
 \centering\epsfig{file=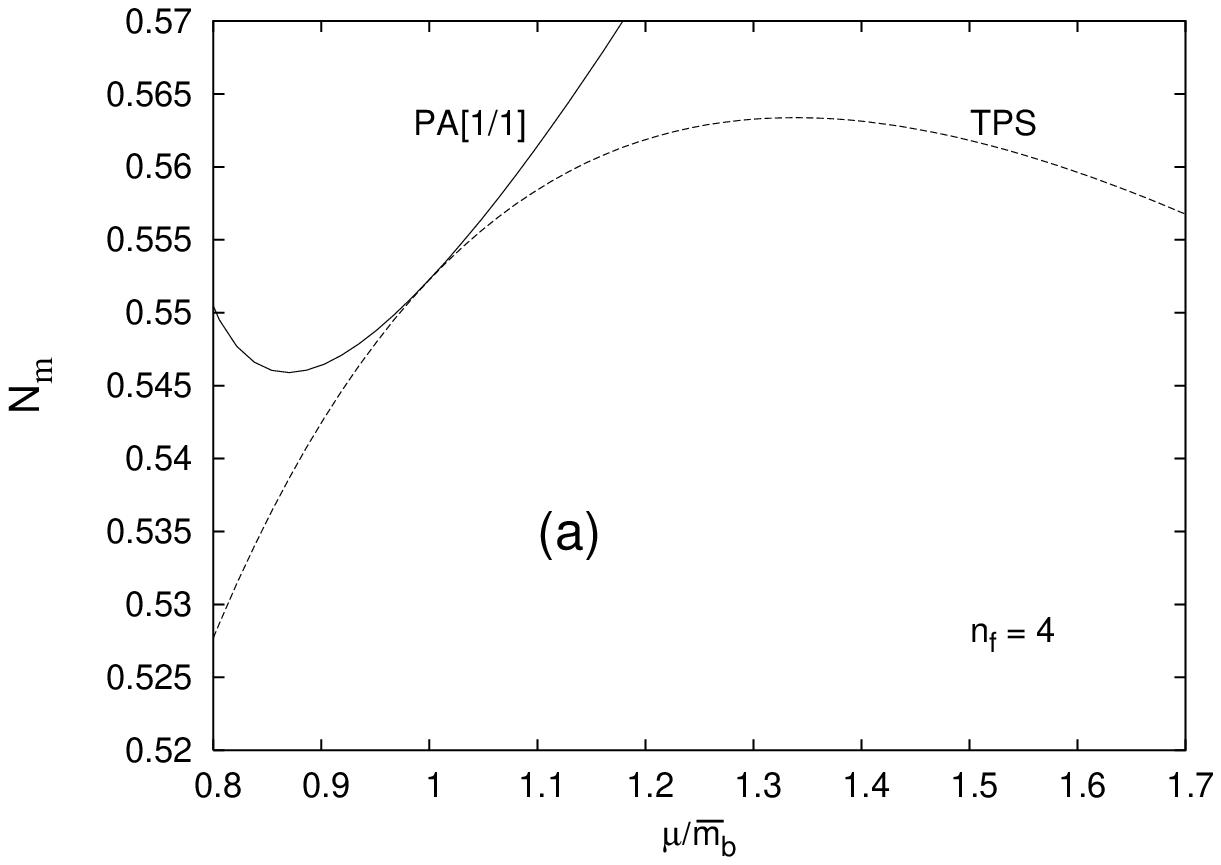,width=\linewidth}
\end{minipage}
\begin{minipage}[b]{.49\linewidth}
 \centering\epsfig{file=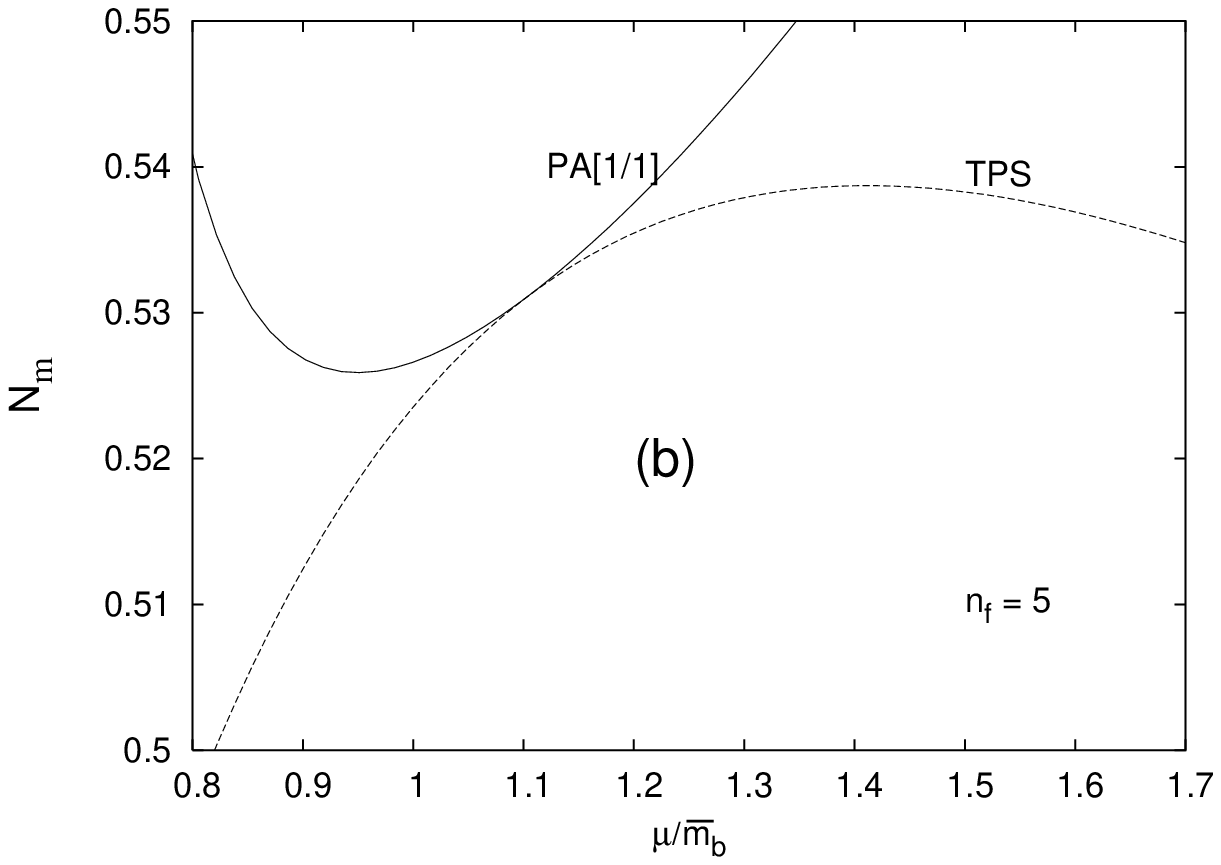,width=\linewidth}
\end{minipage}
\vspace{0.2cm}
\caption{\footnotesize The residue parameter $N_m$ determined
from (${\rm N}^2{\rm LO}$) TPS and Pad\'e $[1/1]$ of 
$R_S(b)$ ($b\!=\!1/2$), as function of the renormalization scale
parameter $x = \mu/{\overline m}_q$, for $n_f=4$ (a) and $n_f=5$ (b).}
\label{Nmvsmu}
\end{figure}

The $b=1/2$ renormalon of $m_q$ \cite{Bigi:1994em}
has a residue that has been determined
with a reasonably high precision \cite{Pineda:2001zq,Lee:2003hh},
due to a good convergence behavior of the series for the residue.
First the calculation of this residue will be re-done, 
focusing on the resulting uncertainties. 
The perturbation expansion for the ratio $S = m_q/{\overline m}_q - 1$,
where ${\overline m}_q = {\overline m}_q(\mu\!=\!{\overline m}_q)$ is the
${\overline {\rm MS}}$ renormalon free mass, is known to order
$\sim\!\alpha_s^3$
\cite{Gray:1990yh}
\begin{eqnarray}
S \equiv  \frac{m_q}{{\overline m}_q} - 1 
&=& \frac{4}{3} a(\mu)
{\Big \{} 1 + a(\mu) \left[ \kappa_1 + \beta_0 L_m \right] 
+ a(\mu)^2 \left[ \kappa_2 + ( 2 \kappa_1 \beta_0 + \beta_1) L_m 
+ \beta_0^2 L_m^2 \right] + {\cal O}(a^3) {\Big \}} \ ,
\label{Smexp}
\\
(4/3) \kappa_1 & = & 6.248 \beta_0 - 3.739 \ ,
\label{k1}
\\
(4/3) \kappa_2 &= &   23.497 \beta_0^2 + 6.248 \beta_1 
+ 1.019 \beta_0 - 29.94 \ ,
\label{k2}
\end{eqnarray}
and $L_m = \ln(\mu^2/{\overline m}_q^2)$. 
The Borel transform is thus known to order $\sim\!b^2$
\begin{equation}
B_{S}(b; \mu) = \frac{4}{3} \left[ 1 + \frac{r_1(\mu)}{1! \beta_0} b +
\frac{r_2}{2! \beta_0^2} b^2 + {\cal O}(b^3) \right] \ ,
\label{BSm1}
\end{equation}
where $r_1 = (\kappa_1 + \beta_0 L_m)$ and $r_2$ are the NLO 
and NNLO coefficients in the expansion (\ref{Smexp}), respectively.
On the other hand, this transform can be rewritten as
\begin{equation}
B_{S}(b; \mu) = \frac{\mu}{{\overline m}_q} 
\frac{\pi N_m }{(1 - 2 b)^{1 + \nu}}\left[ 1 + {\cal O}(1 - 2 b) \right]
+ ({\rm analytic \ term}) \ ,
\label{BSm2}
\end{equation}
where $\nu = \beta_1/(2 \beta_0^2)$ is the power of the
leading ($b = 1/2$) renormalon singularity, the term in the brackets
is independent of $\mu$, and the last term is a function analytic
in the disk $|b| < 1$ (the other renormalons are at $b = -1, -2, \ldots;
+3/2, +2, \ldots$, by Refs.~\cite{Beneke:1994sw,Beneke:1999ui}). 
The $\mu$-independent residue parameter $N_m$ 
can be obtained by constructing the expansion of the following function 
in the powers of $b$:
\begin{equation}
R_S(b; \mu) \equiv  \frac{{\overline m}_q}{\mu} \frac{1}{\pi}
(1 - 2 b)^{1 + \nu} B_{S}(b; \mu) \ .
\label{RSm}
\end{equation}
This function has no pole at $b = 1/2$, only a cut at $b \geq 1/2$
which is a softer singularity. The knowledge of expansion
(\ref{BSm1}) up to $\sim\!b^2$ results in the knowledge of
expansion of $R_S$ up to $\sim\!b^2$. In analogy with an observation
in Refs.~\cite{Lee:1996yk}, the residue parameter $N_m$ is
\begin{equation}
N_m = R_S(b\!=\!1/2) \ ,
\label{Nm1}
\end{equation}
and can thus be obtained approximately from the mentioned
quadratic polynomial (truncated perturbation series - TPS) of $R_S$
in $b$. The convergence of the TPS is good for $\mu \sim {\overline m}_q$.
There is some (unphysical) variation of the TPS predictions when
$\mu$ is varied. Further, if the Pad\'e resummation $[1/1]_{R_S}(b)$
is made, the variation with $\mu$ is different. The results
(TPS and Pad\'e) for $N_m$, as function of $\mu/{\overline m}_q$,
are given in Figs.~\ref{Nmvsmu}(a), (b) for $n_f = 4, 5$, respectively.
The points of zero $\mu$-sensitivity 
(principle of minimal sensitivity -- PMS) give us in the case $n_f\!=\!4$
predictions $N_m = 0.563, 0.546$, for TPS and Pad\'e, respectively,
and in the case $n_f\!=\!5$ predictions $N_m = 0.539, 0.526$.
As central values we take the arithmetic average between the
TPS and Pad\'e values at respective zero $\mu$-sensitivities:
$N_m = 0.555 (n_f\!=\!4); 0.533 (n_f\!=\!5)$. We could also
include in the analysis the Pad\'e predictions at such
$\mu$ where $b^{\rm Pade}_{\rm pole} = -1$, i.e.,
the theoretical nearest-to-origin pole of $R_S(b)$.
However, interestingly, such $\mu$'s are close to the
Pad\'e-PMS $\mu$'s ($\mu/{\overline m}_q \approx 0.85-0.87$ 
when $n_f\!=\!4$; $\mu/{\overline m}_q \approx 0.90-0.95$ 
when $n_f\!=\!5$), and the predictions 
for $N_m$ are virtually the same. The TPS predictions for $N_m$
start falling down fast when $\mu/{\overline m}_q$ decreases below the
value one. However, $\mu/{\overline m}_q < 1$ when the Pad\'e 
approximant gives the value $b_{\rm pole} = -1$
($\mu/{\overline m}_q = 0.85, 0.90$ for $n_f = 4, 5$, respectively).
The absolute value of the uncertainty will be estimated as the deviation
of the TPS prediction of $N_m$ at such low $\mu$
from the aforementioned central values ($0.555, 0.533$).
The resulting estimates are thus
\begin{eqnarray}
N_m(n_f\!=\!4) &=& 0.555 \pm 0.020 \ ,
\label{Nmnf4}
\\
 N_m(n_f\!=\!5) &=& 0.533 \pm 0.020 \ .
\label{Nmnf5}
\end{eqnarray}
On the other hand,
the TPS predictions for $N_m$ at $\mu/{\overline m}_q = 1$
agree exactly with the values obtained by Pineda 
\cite{Pineda:2001zq} ($N_m = 0.5523, 0.5235$, for
$n_f = 4 ,5$, respectively), and are well within the limits 
(\ref{Nmnf4})--(\ref{Nmnf5}). Further, in Ref.~\cite{Lee:2003hh}
the values $N_m \approx 0.557 \pm 0.008, 0.530 \pm 0.008$
for $n_f = 4, 5$ were obtained using a method involving a 
conformal mapping, and they are close to the values
(\ref{Nmnf4})--(\ref{Nmnf5}) here, but the estimated uncertainties 
here are larger.

A somewhat analogous procedure is now applied to
the three-loop expression (\ref{Vr}) for the static potential:
Requiring the reproduction of the values of $N_m$
(\ref{Nmnf4})--(\ref{Nmnf5}) by the static potential
(\ref{Vr}) will allow us to
estimate the values of the unknown three-loop coefficient
expression ${\widetilde a}_3$ appearing in expansion (\ref{Vr})
\begin{equation}
{\widetilde a}_3 \equiv  \frac{a_3}{4^3} + 
\frac{9 \pi^2}{4} \ln \left( \mu_f r e^{\gamma_{\rm E}} \right) 
\label{ta3def}
\end{equation}
as a function of the renormalization scale parameter $\lambda = \mu r$
(it is to be recalled that $b_3 = 9 \pi^2/8$).
Defining the dimensionless quantity $F(r) \equiv (-3/(4 \pi)) r V(r)$,
the TPS (\ref{Vr}) gives us the Borel transform of $F(r)$
up to order $b^3$
\begin{equation}
B_F(b; \mu) = 1 + \frac{v_1}{1! \beta_0} b + 
\frac{v_2}{2! \beta_0^2} b^2 + \frac{v_3}{3! \beta_0^3} b^3 + \cdots \ ,
\label{BFv1}
\end{equation}
where $v_j$ are the coefficients at powers of $a(\mu)$ in expansion
(\ref{Vr})  ($v_1 = a_1/4 + 2 \beta_0 {\ell}$, etc.).
This function has the renormalons at 
$b = 1/2, 3/2, 5/2$, etc.~\cite{Aglietti:1995tg}. It can be written as 
\begin{equation}
B_F(b; \mu r) = \frac{3}{2} \ \mu r \ \frac{N_m}{ (1 - 2 b)^{1 + \nu}}
\left[ 1 + {\cal O}(1 - 2 b) \right]
+ ({\rm analytic \ term}) \ ,
\label{BFv2}
\end{equation}
where $\nu = \beta_1/(2 \beta_0^2)$,
the expression in the brackets is $\mu$-independent,
and the last term is analytic for $|b| < 3/2$.
The parameter $N_m$ here is the same as in expression (\ref{BSm2})
to ensure the leading renormalon cancellation in the Borel transform
of the sum $(V(r)\!+\!2 m_q)$. The function
\begin{equation}
R_F(b; \mu r) \equiv  \frac{2}{3} \frac{1}{\mu r} 
(1 - 2 b)^{1 + \nu} B_{F}(b; \mu r) \ ,
\label{RFv}
\end{equation}
is then less singular at $b=1/2$ (cut instead of pole), and
its evaluation at $b=1/2$ should give us
\begin{equation}
R_F(b\!=\!1/2; \mu r) = N_m \ .
\label{Nmv}
\end{equation}
The coefficients $v_1$ and $v_2$ appearing in
expansions (\ref{Vr}) and (\ref{BFv1}) are explicitly known,
but $v_3$ contains the undetermined three-loop parameter
${\widetilde a}_3$ (\ref{ta3def}) [$\Leftrightarrow \ a_3$]. 
This implies that
the coefficients in the expansion of $R_F(b)$ are known
explicitly up to terms $b^2$, but the coefficient at $b^3$
contains ${\widetilde a}_3$. Thus, any resummation of the
${\rm N}^3 {\rm LO}$ TPS of $R_F(b)$ (i.e., the
TPS including the $b^3$-term) at $b=1/2$, and requiring
the identity (\ref{Nmv}), with $N_m$ values (\ref{Nmnf4})--(\ref{Nmnf5}),
gives us in principle an (approximate) value of the
unknown three-loop coefficient ${\widetilde a}_3$.

From the purely perturbative point of view, the procedure
indicates inconsistency because we are matching the
residue parameter obtained on the basis of the
${\rm N}^2{\rm LO}$ TPS for $R_S(b)$ (\ref{RSm}) with the
residue parameter obtained on the basis of the 
${\rm N}^3 {\rm LO}$ TPS for $R_F(b)$ (\ref{RFv}).
However, in practical terms, there appears to be no
inconsistency. The reason lies in the considerably better 
convergence behavior of the ${\rm N}^2 {\rm LO}$ TPS of $R_S(b)$  
than the ${\rm N}^2 {\rm LO}$ TPS of $R_F(b)$ at $b=1/2$.
The smallness of the term $\sim\!b^2$ in $R_S(b)$
does not seem to be accidental, because this term
is small for any relevant value of $n_f$ ($n_f \leq 5$).
Furthermore, for these reasons,
the procedure of estimating ${\widetilde a}_3$
is indeed rather stable, because 
the ${\rm N}^3 {\rm LO}$ term ($\sim\!b^3$) in $R_F(b)$ 
now plays an important role in ensuring numerically
the $b=1/2$ renormalon cancellation condition (\ref{Nmv}).

\begin{figure}[htb]
\begin{minipage}[b]{.49\linewidth}
 \centering\epsfig{file=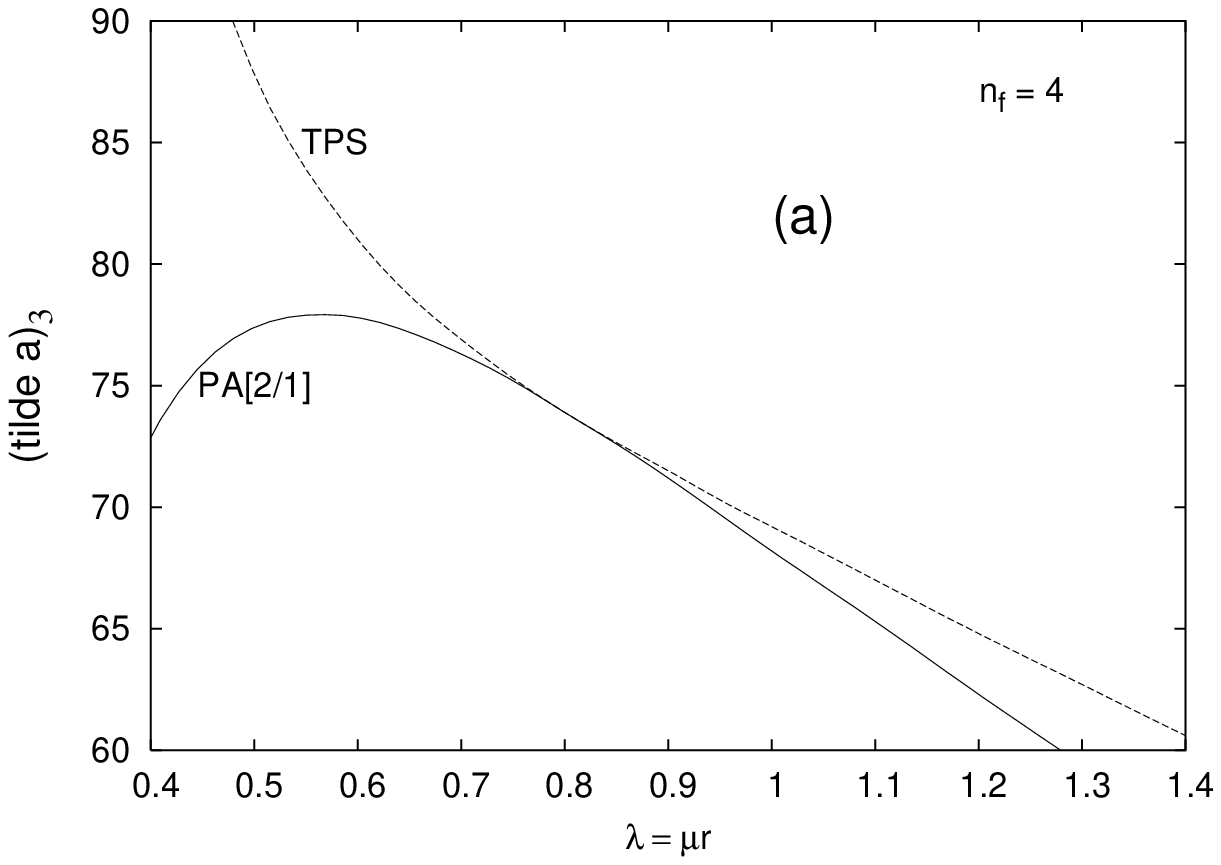,width=\linewidth}
\end{minipage}
\begin{minipage}[b]{.49\linewidth}
 \centering\epsfig{file=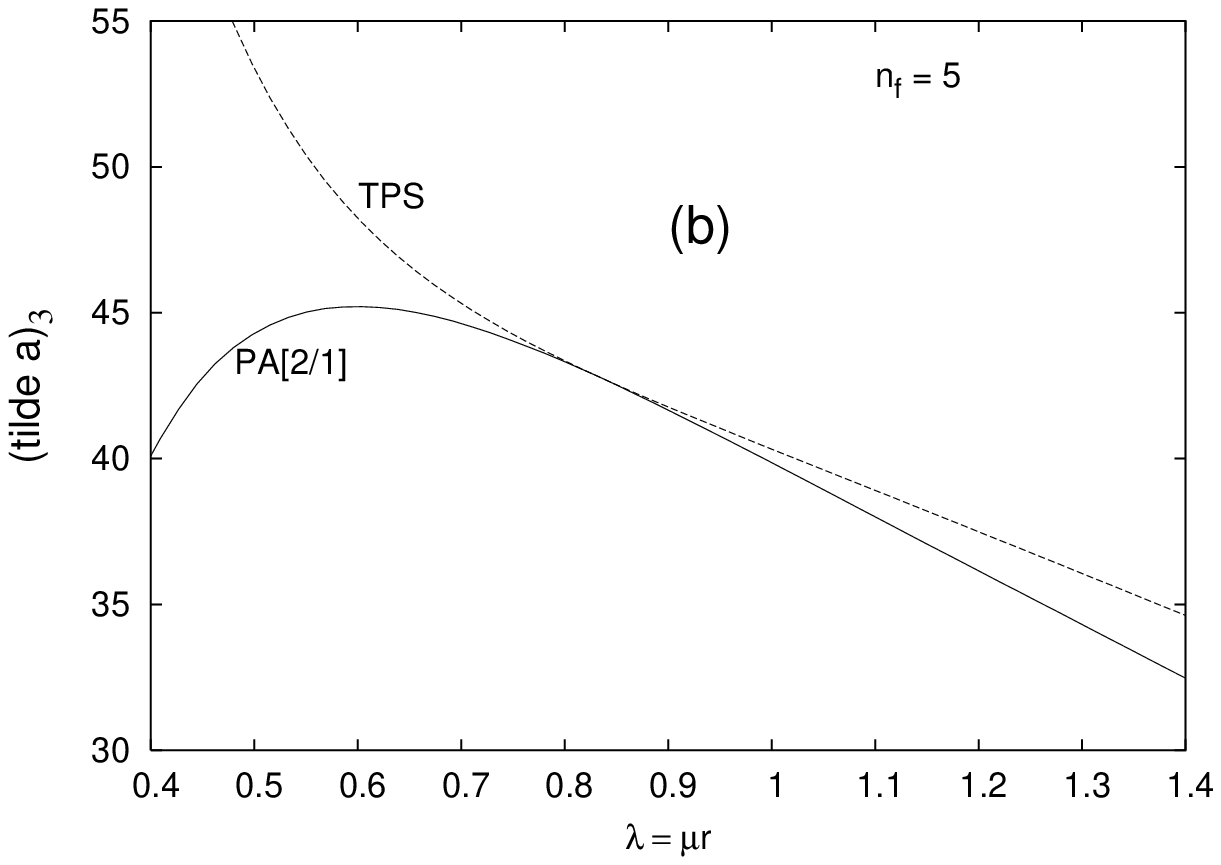,width=\linewidth}
\end{minipage}
\vspace{0.2cm}
\caption{\footnotesize The three-loop parameter
${\widetilde a}_3$ determined from ${\rm N}^3{\rm LO}$ TPS
and Pad\'e $[2/1]$ of $R_F(b)$ ($b\!=\!1/2$) and
requiring the central values (\ref{Nmnf4})--(\ref{Nmnf5}) for
the residue parameter $N_m$, as function of the
renormalization scale parameter $\lambda = \mu r$,
for $n_f=4$ (a) and $n_f=5$ (b).}
\label{ta3vsmu}
\end{figure}

On the other hand, since $R_F(b)$ is a significantly
less singular function than $B_F(b)$ and at the same
time the TPS of $R_F(b)$ shows only slow convergence (at $b=1/2$),
it is very reasonable to apply some form of resummation
beyond the simple TPS evaluation -- e.g., Pad\'e resummation.
The most stable results for ${\widetilde a}_3$, when $\mu$
is varied, are obtained by using the $[2/1](b)$ Pad\'e
of (the ${\rm N}^3 {\rm LO}$ TPS of) $R_F(b)$.
The resulting values of the parameter ${\widetilde a}_3$,
as function of the renormalization scale parameter
$\lambda = \mu r$ and requiring the central values
$N_m$ of Eqs.~(\ref{Nmnf4})--(\ref{Nmnf5}), are presented
in Figs.~\ref{ta3vsmu}(a),(b), for the cases $n_f = 4, 5$,
respectively. 
When Pad\'e $[2/1]$ resummation is applied
to $R_F(b)$, the PMS points do exist, with the values
${\widetilde a}_3 = 77.9, 45.2$ at $\lambda_{\rm PMS} \approx 0.56, 0.60$, 
when $n_f = 4, 5$, respectively. If the values of $N_m$ increase, 
so does ${\widetilde a}_3$. The final result is
\begin{eqnarray}
{\widetilde a}_3(n_f\!=\!4) & = & 77.9^{\ + \ 9.5}_{\ -10.1}(\delta N_m) 
\pm 13.2 (\delta \mu) \ ,
\label{ta3nf4}
\\
{\widetilde a}_3(n_f\!=\!5) & = & 45.2^{\ + \ 7.7}_{\ - \ 8.3} (\delta N_m)
\pm \ 9.1 (\delta \mu) \ .
\label{ta3nf5} 
\end{eqnarray}
where the first uncertainty is due to the uncertainty $\pm 0.020$
in the values of $N_m$ (\ref{Nmnf4})--(\ref{Nmnf5}) (Pad\'e $[2/1]$
was evaluated at the unchanged aforementioned values of $\lambda$), 
and the second uncertainty is due to the renormalization scale
dependence. The latter uncertainty was evaluated as plus/minus
the deviation of ${\widetilde a}_3$ from the mean value
when $\lambda \equiv \mu r$ is increased by factor two from its PMS value. 

The results (\ref{ta3nf4})--(\ref{ta3nf5}) for the three-loop
parameter ${\widetilde a}_3$ (\ref{ta3def}) can be extended further,
in order to estimate the three-loop coefficient $a_3$.
This will involve further uncertainties of the S-US 
factorization scale $\mu_f$ and of the typical distance $r$
between the quarks. As emphasized in Ref.~\cite{Brambilla:1999qa},
the considered $q$-${\bar q}$ singlet static potential
is not the quantity defined via the vacuum
expectation value of a static Wilson loop which includes
US contributions, but rather the relevant object for the
dynamics of the $q$-${\bar q}$ pairs with large
but finite mass. The kinetic and the binding energies
of the $q$-${\bar q}$ system are US energies. Therefore, 
the interactions of the US gluons with the $q$-${\bar q}$ system 
are sensitive to the energies of that system and
should be excluded from the static potential \cite{Brambilla:1999qa}.
The exclusion of these interactions leads to the
term $\propto \ln(\mu_f r e^{\gamma_{\rm E}})$
at $\sim\!a^4$ in the potential (\ref{Vr}). Hence the 
value of the parameter $a_3$ will depend on the
values of the scales $\mu_f$ and $r$. 
The dependence of $V(r)$ on the IR cutoff $\mu_f$
of the soft scale regime,
and the consequent dependence of the values of
$N_m$ (\ref{Nmv}) and $a_3$ [cf.~Eq.~(\ref{ta3def})]
on $\mu_f$, does not imply that US contributions affect them,
but rather that the US contributions to
those quantities are cut out, zero. 

Since the scales $\mu_f$ and $r$ appear in
logarithms, a natural choice for $\mu_f$ would be 
$\mu_f = (E_{\rm S} E_{\rm US})^{1/2}$.
The typical S scale is of the order of the
three-momentum transfer $|{\bf k}|$ in the $q$-${\bar q}$
ground state $|1\rangle$: $E_{\rm S} \sim m_q \alpha_s$.
The typical US scale is 
$E_{\rm US} \sim |E_{q {\bar q}}| \sim m_q \alpha_s^2$.
Therefore, the factorization scale can be estimated as
\begin{equation}
\mu_f \left[ \approx (E_{\rm S} E_{\rm US})^{1/2} \right] = \kappa \ m_q 
\alpha_s(\mu_s)^{3/2} \ ,
\label{muf}
\end{equation}
where $\kappa \sim 1$ and we take $\mu_s \approx \mu$ ($\sim\!E_{\rm S}$).
The typical distance $r$ in the combination
${\widetilde a}_3$ (\ref{ta3def}) can be estimated in the
following way. First the expectation value of the
static potential term proportional to ${\widetilde a}_3$
is calculated in the $q$-${\bar q}$ ground state $|1 \rangle$
\begin{eqnarray}
\langle 1 | \frac{1}{r} a^4(\mu) {\widetilde a}_3(r) | 1 \rangle &=&
\frac{1}{{\widetilde a}_{\rm B}(\mu)} a^4(\mu) \left[ \frac{a_3}{4^3} +
\frac{9 \pi^2}{4} \ln 
\left( \mu_f {\widetilde a}_{\rm B}(\mu) e/2 \right) \right]
\label{EV1}
\\
& = & \frac{1}{{\widetilde a}_{\rm B}(\mu)} a^4(\mu) 
\left[ \frac{a_3}{4^3} +
\frac{9 \pi^2}{4} \ln \left( \frac{3 e}{4} \frac{\kappa}{{\widetilde \kappa}}
\alpha^{1/2}(\mu_s) \right)
\right] \ .
\label{EV2}
\end{eqnarray}
Here, ${\widetilde a}_B(\mu) = 3/(2 m_q {\widetilde \alpha}_s(\mu))$ is the 
modified Bohr radius \cite{Titard:1993nn} appearing in the $q{\bar q}$ 
ground state wavefunction, and we denoted
\begin{eqnarray}
 {\widetilde \kappa} =  \frac{{\widetilde \alpha}_s(\mu)}{\alpha_s(\mu)} &=&
1 + a(\mu) \left[ \frac{1}{4} a_1 + 2 \beta_0 \gamma_{\rm E} \right]
+ a(\mu)^2 \left[ \frac{1}{16} a_2 + 
\left( a_1 \beta_0 + 2 \beta_1 \right) \gamma_{\rm E} +
\beta_0^2 \left( 4 \gamma_{\rm E}^2 + \pi^2/3 \right) \right] + \cdots \ .
\label{tildekappa}
\end{eqnarray}
In Eq.~(\ref{EV2}), the
expression (\ref{muf}) for $\mu_f$ was inserted.
Eqs.~(\ref{EV1})--(\ref{EV2}) suggest that we can estimate the value of $a_3$
from the value of ${\widetilde a}_3$ via the relation
\begin{equation}
\frac{1}{4^3} \ a_3 \approx {\widetilde a}_3 - 
\frac{9 \pi^2}{4} \ln \left( 2.04 \ \frac{\kappa}{{\widetilde \kappa}} 
\alpha_s^{1/2}(\mu_s) \right) \ .
\label{a3v1}
\end{equation}
Now we estimate the values of $\kappa$ and ${\widetilde \kappa}$
for the  $b$-${\bar b}$ and $t$-${\bar t}$ system.

Since $\mu_s \sim m_q \alpha_s(\mu_s)$, we have
for the $b$-${\bar b}$ system ($m_q \approx 5.$ GeV, $n_f = n_l = 4$), 
the values $\mu_s \approx 1$-$2$ GeV and $\alpha_s(\mu_s) \approx 0.3$
(as suggested by resummations of the semihadronic $\tau$-decay
width \cite{Cvetic:2001sn}). The series (\ref{tildekappa}) is
strongly divergent then $(1 + 0.37 + 0.40 + \cdots)$, 
and we can estimate ${\widetilde \kappa}(n_f\!=\!4) \approx 1.25 \pm 0.10$
by applying the resummation method of \cite{Lee:2002sn}.\footnote{
Note that the series (\ref{tildekappa}) represents 
the quantity $(-3/(4 \pi)) r V(r)$ of Eq.~(\ref{Vr}) at 
$\mu r = 1$. The principal value (PV) prescription was taken for
the Borel integration over the $b=1/2$ renormalon singularity.}
The resummed binding energy in the ground
state is $E_{b {\bar b}} \approx - 0.4$ GeV 
\cite{Contreras:2003zb,Lee:2003hh},
i.e., $E_{\rm US} \approx |E_{b {\bar b}}| \approx 
0.9 \ m_b \alpha_s(\mu_s)^2$.
On the other hand, $E_{\rm S}$ is approximately equal to the
expectation value of the three-momentum in the ground state
\begin{equation}
E_{\rm S} \approx \langle 1 | \ |{\bf p}| \ |1 \rangle =
\frac{8}{3 \pi} \frac{1}{{\widetilde a}_B(\mu)} = 
\frac{16}{9 \pi} {\widetilde \kappa} m_q \alpha_s(\mu) \ .
\label{ES}
\end{equation}
Therefore, Eq.~(\ref{muf}) gives 
$\kappa \approx 0.7 {\widetilde \kappa}^{1/2}$ 
($\approx 0.8$) when $n_f\!=\!4$. 

In the case of the $t$-${\bar t}$ system
($m_q \approx 175$ GeV, $n_f = n_l = 5$),
we have $\mu_s \approx 30$ GeV, and $\alpha_s(\mu_s) \approx 0.14$.
The series (\ref{tildekappa}) is less divergent ($1 + 0.15 + 0.07 + \cdots)$,
and we estimate with the aforementioned resummation
${\widetilde \kappa}(n_f\!=\!5) \approx 1.30 \pm 0.10$.
The resummed binding energy in the ground
state is $E_{t {\bar t}} \approx - 3.$ GeV 
\cite{Contreras:2003zb,Lee:2003hh},
i.e., $E_{\rm US} \approx |E_{t {\bar t}}| \approx 
0.9 \ m_t \alpha_s(\mu_s)^2$. 
Eqs.~(\ref{muf}) and (\ref{ES}) then give  
$\kappa \approx 0.7 {\widetilde \kappa}^{1/2}$ 
($\approx 0.8$) when $n_f\!=\!5$.    

Therefore, the approximate relation (\ref{a3v1}) can then be rewritten
explicitly for the $b$-${\bar b}$ and $t$-${\bar t}$ system as
\begin{equation}
\frac{1}{4^3} \ a_3 \approx  {\widetilde a}_3 - 
\frac{9 \pi^2}{4} \ln \left( 1.4 \ \frac{1}{{\widetilde \kappa}^{1/2}} 
\alpha_s^{1/2}(\mu_s) \right) \ ,
\label{a3vnf}
\end{equation}
where, as mentioned, we take the central values:
$\alpha_s(\mu_s) \approx 0.3$ and ${\widetilde \kappa} \approx 1.25$ 
for $n_f\!=\!4$;
$\alpha_s(\mu_s) \approx 0.14$ and ${\widetilde \kappa} \approx 1.3$ 
for $n_f\!=\!5$.
This then gives
\begin{eqnarray}
\frac{1}{4^3} \ a_3(n_f\!=\!4) & = & 
86.3^{\ + \ 9.5}_{\ - 10.1} (\delta N_m) \pm 13.2 (\delta \mu) 
^{ \ + \ 15.4}_{\ - \ 8.4} (\delta \mu_f) \ ,
\label{a3nf4}
\\
\frac{1}{4^3} \ a_3(n_f\!=\!5) & = & 
62.5^{\ + \ 7.7}_{\ - \ \ 8.3} (\delta N_m) \pm 9.1 (\delta \mu) 
\pm  15.4 (\delta \mu_f) \ ,
\label{a3nf5}
\end{eqnarray}
where the last uncertainty is from the
uncertainty in the argument of the logarithm in Eq.~(\ref{a3vnf})  
involving the estimates of the soft and ultrasoft scales. 
These uncertainties are large, because
the central values in the logarithm of Eq.~(\ref{a3vnf})
are only suggestive. The value $\pm 15.4$ for these uncertainties
was obtained by allowing the overall factor $2$ or $1/2$ in
the logarithm in Eq.~(\ref{a3vnf}). On the other hand, the
smaller downward uncertainty $-8.4$ in the case $n_f\!=\!4$
is due to the problematically small hierarchy between the
soft and ultrasoft scales in this case, and the related
minimal requirement that the factorization scale
$\mu_f$ should be below the typical soft scale $\sim\!|{\bf k}|$.
Stated otherwise, the logarithmic term in Eq.~(\ref{ta3def})
should be negative, or equivalently, the IR cutoff term
proportional to $b_3$ in expansion (\ref{Vr}) should give
a positive contribution to $V(r)$ (i.e., $\mu_f < 0.56/r$).

We now investigate how numerically important
the obtained $a_3$-terms are with respect to other terms
in the binding energy of the quarkonium. 
For example, in the $b {\bar b}$ ground state
$\Upsilon(1S)$, the typical value of the interquark distance
is $r \sim 1/(m_b \alpha_s(\mu_s))$, i.e., 
$r \approx 0.50$-$0.75 \ {\rm GeV}^{-1}$. Table
\ref{table1} shows the values of the pure $a_3$-term
of $V(r)$ for such $r$, for the central $a_3$ value
(\ref{a3nf4}), as well as the values of the entire leading order
(LO; $\sim a/r$), NLO ($\sim a^2/r$), 
${\rm N}^2{\rm LO}$ ($\sim a^3/r$),
and ${\rm N}^3{\rm LO}$ ($\sim a^4/r$) parts of $V(r)$ of Eq.~(\ref{Vr}).
\begin{table}
\caption{\label{table1} The separate contributions of the
perturbative potential $V(r)$ (\ref{Vr}) for two
typical radia $r$ of the $b {\bar b}$ system ($n_f\!=\!4$).
The $a_3$-contribution is given for the central value 
(\ref{a3nf4}) of $a_3$. The values are given for 
the renormalization scale $\mu = 2/r$, and for 
$\mu = 1/r$ in parentheses. Included
are also values of the logarithmic confining potential
of Ref.~\cite{Brisudova:1996vw}. All energies are in MeV;
radia are in ${\rm GeV}^{-1}$. Other details are 
given in the text.}
\begin{ruledtabular}
\begin{tabular}{cccccccc}
$r$ & $V_{\rm conf}$ & $V(a_3)$ & $V({\rm LO})$ & 
$V({\rm NLO})$ & $V({\rm N}^2{\rm LO})$ & $V({\rm N}^3{\rm LO})$ & 
$\alpha_s(\mu)$ \\
\hline
$0.50$ & $+19$ & $-23 ( \ -74)$ & $-629 (-842)$ & $-319 (-328)$ &
$ -281 (-374)$  &$-326 (-553)$ &
$0.236 (0.316)$
\\
$0.75$ & $+43$ & $-29 (-126)$ & $-491 (-710)$ & $-292 (-350)$ &
$-301 (-505)$  &$-409 (-944)$ &
$0.276 (0.399)$  
\\
\end{tabular}
\end{ruledtabular}
\end{table}
These values are given for two choices of the renormalization
scale: $\mu = 2/r, 1/r$. 
The values of the $a_3$-term as given in the Table
should be regarded with some caution though, because they strongly
depend on the choice of $\mu$ and because the
perturbative series for $V(r)$ is asymptotically
divergent. This can be seen from the values
of different orders of $V(r)$ given in the Table.
Borel-related resummation methods for the binding energy $E_{b \bar b}$
\cite{Contreras:2003zb} give for the $a_3$-term
contribution $(- 16 \pm 4)$ MeV,
i.e., $E_{b {\bar b}}(a_3) - E_{b {\bar b}}(0) \approx (-16 \pm 4)$ MeV,
when the values of $a_3$ are taken from Eq.~(\ref{a3nf4})\footnote{
The uncertainties were added in quadrature: $a_3/4^3 \approx 86 \pm 23$.} 
and $\mu \sim E_{\rm S}$. In addition, we display in Table \ref{table1} the
values of the logarithmic confining potential 
$V_{\rm conf}(r) = (C_F \alpha {\cal L}/\pi) V_0({\cal L}r)$
of Ref.~\cite{Brisudova:1996vw}, 
based on the renormalization-group-improved
light-front Hamiltonian formalism of 
Refs.~\cite{Wilson:1994fk,Brisudova:1995hv}
\begin{eqnarray}
V_{\rm conf}(r) & = & \frac{4}{3 \pi} \alpha {\cal L}
\left[ 2 \ln R - 2 {\rm Ci}(R) + \frac{4}{R} {\rm Si}(R) - \frac{2}{R^2}
(1\!-\!\cos R ) + \frac{2}{R} \sin R - 5\!+\!2 \gamma_{\rm E} \right] ,
\label{Vconf}
 \end{eqnarray}
where: $R = {\cal L} r$; ${\cal L}$ is a cutoff scale parameter
of the framework; $\alpha$ is an effective $\alpha_s$.
${\rm Ci}$ and ${\rm Si}$ are the usual cosine and sine integrals:
${\rm Ci}(z) = - \int_z^{\infty} [\cos(t)/t] dt$;
${\rm Si}(z) = \int_0^{z} [\sin(t)/t] dt$.
The fitted values of the parameters in that potential are
also taken from Ref.~\cite{Brisudova:1996vw}: 
$c\!=\!1$ [$c \equiv (4/3) (m_b \alpha/{\cal L})$],
$m_b\!=\!4.8$ GeV, $\alpha\!=\!0.4$
($\Rightarrow {\cal L}\!=\!2.56$ GeV).
The nonperturbative contribution to the binding
energy $E_{b \bar b}$ can alternatively be estimated as coming from the
gluonic condensate \cite{Voloshin:hc}
\begin{equation}
E_{b \bar b}(us)^{\rm (np)} \approx 
{\overline m}_b 
\pi^2 \frac{624}{425} 
\left( \frac{4}{3} {\overline m}_b \alpha_s(\mu_{us}) \right)^{-4} 
{\big \langle} a(\mu_{us}) G_{\mu \nu} G^{\mu \nu} {\big \rangle}
\approx  (50 \pm 35) \ {\rm MeV} \ ,
\label{Ebbusnp}
\end{equation}
where ${\overline m}_b = 4.2$ GeV was taken,
and the ultrasoft energy $\mu_{us} \sim m_b \alpha_s^2 < 1$ GeV
was taken equal to $\mu \approx 1.5$-$2.0$ GeV
in order to be able to determine $\alpha_s(\mu_{us})$
still perturbatively [$\alpha_s(\mu) \approx 0.30$-$0.35$].
Thus, the confining energies of Table \ref{table1}
and of Eq.~(\ref{Ebbusnp}) consistently give
nonperturbative contributions to $E_{b \bar b}$
roughly in the range $20$-$50$ MeV, compared to 
$(-16 \pm 4)$ MeV of the $a_3$-term. We see that the $a_3$-term
in $b {\bar b}$ gives a nonnegligible contribution
in comparison to the nonperturbative contributions
to the binding energy (and thus to the meson mass).
On the other hand, for the lighter charmonium the 
nonperturbative effects are expected to be more important, 
while for the much heavier toponium less important than 
the $a_3$-term contributions. 

The contributions to $E_{b \bar b}$ from the ``hard'' ($\sim m_b$)
and ``soft'' ($\sim m_b \alpha_s$) energy regimes give about 
$-300$ MeV, and those from the 
ultrasoft regime ($\sim m_b \alpha_s^2$) about $-150$ MeV
\cite{Contreras:2003zb}.
By the virial theorem, this implies for the kinetic energy an estimate
of $150$-$250$ MeV.

We now compare the results (\ref{ta3nf4})--(\ref{ta3nf5}) and 
(\ref{a3nf4})--(\ref{a3nf5})
with those in the literature.
The authors of Ref.~\cite{Chishtie:2001mf}
have estimated the three-loop coefficient of the
static potential $V(|{\bf k}|)$ with a Pad\'e-method based on the
known one- and two-loop coefficients.
However, they did not have the IR cutoff term in the three-loop 
coefficient. Therefore, there is no
direct correspondence with the results presented here, but
in $V(r)$ their coefficient $c_0$ formally may correspond
to ${\widetilde a}_3$ of Eq.~(\ref{ta3def}),
or equivalently, to $a_3/4^3$ if $\mu_f = \exp(-\gamma_{\rm E})/r$.
Their estimated values were $c_0 = 97.5, 60.1$ for
$n_f=4,5$, which are higher than the values (\ref{ta3nf4})--(\ref{ta3nf5}).
The estimates of T.~Lee \cite{Lee:2003hh},
also without the effects of the IR cutoff,
and based on a Borel transform method \cite{Jeong:2002ph}, give
$a_3/4^3 = 59 \pm 81 (n_f\!=\!4)$; $34 \pm 63 (n_f\!=\!5)$.
Further, Pineda \cite{Pineda:2001zq} estimated the three-loop
coefficient of $V(r)$ from renormalon-dominated large order
behavior of the coefficients for
$\mu = 1/r$ but without consideration of the IR cutoff
effects; if his results are to be interpreted with the value of the
factorization scale $\mu_f = 1/r$
($\mu_f \equiv \nu_{us}$ in the notation of \cite{Pineda:2001zq}),
as suggested by him, then this would give, in the notation 
(\ref{ta3def}) of the present paper, the values
$a_3/4^3 =  59.6, 24.3$ for $n_f = 4, 5$. On the other hand,
in terms of ${\widetilde a}_3$ of Eq.~(\ref{ta3def}), his
results would imply ${\widetilde a}_3 = 72.4, 37.1$
for $n_f = 4, 5$, which are not very far from the values 
(\ref{ta3nf4})--(\ref{ta3nf5}).
Pineda, in the framework of his method, did account for the leading 
IR renormalon cancellation in $(V(r)\!+\!2 m_q)$.

The main results and conclusions of the present work are the following:
The unknown part $a_3$ of the three-loop coefficient of the
static potential (\ref{Vr}) was estimated by using the
known property that the leading infrared renormalon
singularities cancel in the sum $V(r) + 2 m_q$. Further, the 
presented method takes into account the fact that the contributions of the 
ultrasoft energy regime should be excluded from the analysis. 
The obtained estimated values of $a_3$ for the $b {\bar b}$ and
$t {\bar t}$ quarkonia are given in Eqs.~(\ref{a3nf4})
and (\ref{a3nf5}), respectively.  
In the $b {\bar b}$ system, the obtained value of 
the three-loop parameter $a_3$ leads to a decrease of the 
binding energy by about $10$-$20$ MeV. This is smaller than, but still
comparable to, the increase of the binding energy by about $20$-$50$ MeV
by nonperturbative effects. The similar value (\ref{a3nf5})
of $a_3$ for the much heavier $t {\bar t}$ system is expected to influence 
the binding energies to a similar degree as in $b {\bar b}$, but
the nonperturbative effects are expected to be much weaker. 
For the lighter $c{\bar c}$ system the nonperturbative
effects are expected to dominate over the three-loop effects.

\begin{acknowledgments}
The author thanks to A.A.~Penin for several clarifications, 
and to C.~Contreras and P.~Gaete for useful discussions.
This work was supported by FONDECYT (Chile) 
grant No. 1010094.
\end{acknowledgments}

\end{document}